\documentclass[5p]{elsarticle}
\makeatletter
\def\ps@pprintTitle{%
 \let\@oddhead\@empty
 \let\@evenhead\@empty
 \def\@oddfoot{\hfill{\thepage}\hfill}%
 \let\@evenfoot\@oddfoot}
\makeatother

\usepackage{hyperref}
\usepackage[utf8x]{inputenc}
\usepackage{amsmath}
\usepackage{amssymb}
\usepackage{bbm}

\newcommand{\id}{\mathbf{1}}
\newcommand{\un}{\mathbbm{1}}
\newcommand{\act}{\triangleright}

\begin{document}

\begin{frontmatter}

\title{Covariant four dimensional differential calculus in $\kappa$-Minkowski}

\author{Giacomo Rosati}
\ead{giacomo.rosati@uwr.edu.pl}
\address{Institute for Theoretical Physics, University of Wroc{\l}aw, Pl. Maksa Borna 9, Pl–50-204 Wroc{\l}aw, Poland}





\begin{abstract}
It is generally believed that it is not possible to have a four dimensional differential calculus in $\kappa$-Minkowski spacetime, with $\kappa$-Poincar\'e relativistic symmetries, covariant under ($\kappa$-deformed) Lorentz transformations. Thus, one usually introduces a fifth differential form, whose physical interpretation is still challenging, and defines a covariant five dimensional calculus. Nevertheless, the four dimensional calculus is at the basis of several works based on $\kappa$-Minkowski/$\kappa$-Poincar\'e framework that led to meaningful insights on its physical interpretation and phenomenological implications.
We here revisit the argument against the covariance of the four dimensional calculus, and find that it depends crucially on an incomplete characterization of Lorentz transformations in this framework. 
In particular, we understand that this is due to a feature, still uncovered at the time, that turns out to be fundamental for the consistency of the relativistic framework:
the noncommutativity of the Lorentz transformation parameters. Once this is taken into account, the four dimensional calculus is found to be fully Lorentz covariant.
The result we obtain extends naturally to the whole $\kappa$-Poincar\'e algebra of transformations, showing the close relation between its relativistic nature and the properties of the differential calculus.

\end{abstract}


\end{frontmatter}


\section{Introduction}

$\kappa$-Poincar\'e algebra ${\cal P}_\kappa$ was introduced in~\cite{Luk91,Luk92} as a possible Hopf deformation of standard special relativistic Poincar\'e symmetries.
In~\cite{MajidRuegg} it was shown that it can be given the structure of a semidirect product of the classical Lorentz group $so(1,3)$ acting in a deformed way on the translation sector $T$, a structure known as bicrossproduct, and denoted as $U(so(1,3)) \triangleright\!\!\!\blacktriangleleft T$. It was shown also in~\cite{MajidRuegg} that one can introduce a noncommutative space of coordinates ${\cal M}_\kappa$ on which ${\cal P}_\kappa$ acts covariantly,  denoted as $\kappa$-Minkowski.
A (first order) differential calculus on ${\cal M}_\kappa$ can be defined according to the Woronowicz formulation of bicovariant differential calculi on quantum groups~\cite{WoroDiffCalc}.
In~\cite{Sitarz5dCalc} (see also~\cite{gonera5dCalc}), Sitarz introduced a notion of ``Lorentz'' covariance for the differential calculus on ${\cal M}_\kappa$ according to which, in order to comply also with bicovariance, one needs to add a fifth differential form, so that the calculus becomes five dimensional.

A further study of bicovariant differential calculi on ${\cal M}_\kappa$ was carried out in~\cite{OecklDiffCalc} (see also~\cite{MajidOeckl}), identifying in particular a proposal for a four dimensional calculus compatible with a notion of translational invariant integration on ${\cal M}_\kappa$.
The four dimensional differential calculus proposed in~\cite{OecklDiffCalc}~\footnote{Actually in~\cite{OecklDiffCalc} two versions of the 4D calculus are proposed, one is used in~\cite{GACmajid}, while the other in~\cite{GACagoDandreakMinkTransl,GACkNoether}.} was at the basis of the definition of Fourier transforms and plane waves on ${\cal M}_\kappa$ in~\cite{GACmajid}, as well as of a description of noncommutative translation parameters~\cite{GACagoDandreakMinkTransl} that led to the first results on Noether analysis for $\kappa$-Minkowski field theories~\cite{GACkNoether}, that contributed towards a physical interpretation of noncommutative theories based on $\kappa$-Poincar\'e symmetries.
These studies are among the premises of the research program of quantum gravity phenomenology based on Planck-scale (Hopf) deformed relativistic symmetries (see for instance~\cite{kbob,IceCubePLB,IceCubeNat}).

However, the results of~\cite{Sitarz5dCalc} seem to rule out the four dimensional calculus introduced in~\cite{OecklDiffCalc}, if one doesn't want to break Lorentz invariance.
It is on this ground that in the relevant literature it is asserted that in $\kappa$-Minkowski spacetime there is no four dimensional calculus which is (Lorentz) covariant~\footnote{In particular in~\cite{JuricMeljakDiff} (see also references therein for the previous works of the authors) a classification of four dimensional bicovariant differential calculi for spacetimes with noncommutativity of Lie algebra type, that include $\kappa$-Minkowski within the ``timelike'' class of deformations, is given. It is concluded, adopting the notion of covariance given in~\cite{Sitarz5dCalc}, that the only Lorentz covariant calculi belong to the class of ``lightlike'' deformations, thus excluding $\kappa$-Minkowski (at least in its original formulation).}\footnote{In~\cite{FlaviokDiff} a more recent formulation of the 5D differential calculus, with the introduction of other quantum differential geometry notions, is presented.}~\cite{LukKosMasSitkField,AgostinikDiff,KowaNowaFrePLB,KowaNowaFre,FlaviokDiff,JuricMeljakDiff}, so that some authors~\cite{AgostinikDiff,KowaNowaFrePLB,KowaNowaFre} tend to prefer a description of translations for ${\cal M}_\kappa$ noncommutative field theories based on the five dimensional calculus of~\cite{Sitarz5dCalc}.
The question of which differential calculus to use is not just a formal matter though. 
It is well known that among the most interesting frameworks for quantum gravity phenomenology are the ones that depend on the possibility to have a deformed dispersion relation characterizing relativistic kinematics, as for instance in studies of Planck-scale in-vacuo dispersion~\cite{kbob,IceCubePLB,IceCubeNat}, a feature that seems to emerge naturally~\cite{GACmajid} in scenarios based on the four dimensional calculus~\cite{OecklDiffCalc}.
On the contrary, as it was pointed out already in~\cite{AgostinikDiff}, the resulting Noether charges associated to a description of translations based on the 5D calculus~\cite{KowaNowaFrePLB,KowaNowaFre} (see also the recent~\cite{kdiscrete}) satisfy the standard, undeformed, special relativistic algebra, and, if one interprets physical energy and momentum to be the conserved charges arising from Noether analysis, one finds that they satisfy the standard undeformed dispersion relation.

We find that these considerations, mainly the importance of the calculus for a physical interpretation of the noncommutative theory and its implications for Planck-scale phenomenology, motivate to revise the argument put forward in~\cite{Sitarz5dCalc} against the covariance of the 4D calculus, also in light of some more recent results regarding the properties of the $\kappa$-Poincar\'e symmetries.
In particular, it was pointed out in~\cite{GACnopure} that the relativistic description of the whole $\kappa$-Poincar\'e framework requires to introduce, besides noncommutative translation parameters~\cite{GACkNoether}, noncommutative boost and rotation transformation parameters associated to Lorentz sector of the algebra.
We find that when one takes into account of the role of the Lorentz noncommutative parameter, the four dimensional calculus of~\cite{OecklDiffCalc} complies naturally with a notion of covariance under Lorentz transformations.
Moreover, we show that this notion of covariance extends to the whole $\kappa$-Poincar\'e algebra of relativistic transformations, in a way that the Lorentz transformations are also described by a (second) bicovariant 4D differential calculus.
To further motivate the introduction of noncommutative Lorentz parameters, we show how they are necessary for the definition of Lorentz covariant plane waves on ${\cal M}_\kappa$, and thus of scalar fields.
We conclude with a description of a ``pregeometric''~\cite{pregeometry} representation of the calculi, that allows to highlight some of the features of the 
calculi, and we outline some considerations for a comparison with the 5D calculus.

But, first of all, we introduce the notion of Lorentz covariance for a (bicovariant) differential calculus from relativistic considerations.

\section{Notion of covariance}
\label{sec:covariance}

Consider a noncommutative spacetime defined by commutation relations of Lie algebra type
\begin{equation}
[x^\mu,x^\nu] = \gamma^{\mu\nu}_\rho x^\rho \ ,
\label{noncommLie}
\end{equation}
with primitive coproducts
\begin{equation}
\Delta x^\mu = x^\mu \otimes \un + \un + x^\mu \ ,
\label{coproductX}
\end{equation}
and a four dimensional differential calculus on (\ref{noncommLie}), such that
\begin{equation*}
x^\mu \rightarrow x^\mu + dx^\mu \ .
\label{diffCalc}
\end{equation*}
The differential calculus must first of all satisfy the Leibniz rule
\begin{equation}
d(f(x)g(x)) = df(x) g(x) + f(x) dg(x) \ ,
\label{diffCalcLeibniz}
\end{equation}
where it is also assumed that $d^2 =0$.
Moreover, Woronowicz's formulation~\cite{WoroDiffCalc} requires compatibility conditions with the Hopf structures known as bicovariance.
One can show (see for instance~\cite{Sitarz5dCalc,MajidOeckl}) that, due to the simplicity of the coproduct (\ref{coproductX}), the bicovariance of the differential calculus reduces to two conditions: compatibility with the defining commutators (\ref{noncommLie})
\begin{equation*}
d[x^\mu,x^\nu] = [dx^\mu,x^\nu] + [x^\mu,dx^\nu]\ ,
\end{equation*}
and the Jacobi identities
\begin{equation*}
\left[\left[x^{\mu},x^{\nu}\right],dx^{\rho}\right]+\left[\left[x^{\nu},dx^{\rho}\right],x^{\mu}\right]+\left[\left[dx^{\rho},x^{\mu}\right],x^{\nu}\right]=0 \ .\end{equation*}
It is easy to see, assuming the independence of the $dx^\mu$, that these two sets of conditions combine to
\begin{equation*}
[x^\mu + dx^\mu,x^\nu + dx^\nu] = \gamma^{\mu\nu}_\rho (x^\rho + dx^\rho) \ ,
\label{diffCalcIsomorph}
\end{equation*}
i.e. the map $\id + d$ is an algebra isomorphism.

An (infinitesimal) Lorentz transformation (boost plus rotation) is generated by the operator
\begin{equation}
L\act = (\id + i \xi^j N_j + i \rho^j R_j) \act \ ,
\label{Lorentz}
\end{equation}
so that, a (scalar) function of $x^\mu$ changes as
\begin{equation*}
f(x') = L\act f(x) \ .
\end{equation*}
We define the action of Lorentz on the differential, according to~\cite{Sitarz5dCalc}, as given by
\begin{equation}
d{x'}^\mu = L\act dx^\mu = dL\act x^\mu \ .
\label{LorentzActDiff}
\end{equation}
We will actually show in Sec. that this assumption is indeed justified from a relativistical point of view.

We now propose our definition of covariance for the differential calculus, that stems from the following relativistic consideration: since the Lorentz transformation (\ref{Lorentz}) defines the relation between coordinates of events for observers that are relatively boosted or rotated, what we can ask as covariance, amounts to the requirement that the differential calculus satisfies the same properties for both observers. Thus, to be covariant, a differential calculus must satisfy the property
\begin{equation}
[{x'}^\mu + d{x'}^\mu , {x'}^\nu + d{x'}^\nu ] = \gamma^{\mu\nu}_\rho ({x'}^\rho + d{x'}^\rho) \ ,
\label{covariance}
\end{equation}
or, equivalently,
\begin{equation*}
[L\act (x^\mu+dx^\mu), L\act (x^\nu+dx^\nu)] = L\act[x^\mu+dx^\mu, x^\nu+dx^\nu]  \ .
\end{equation*}
Notice that there is a crucial difference between this condition and the one proposed in~\cite{Sitarz5dCalc}, which was~\footnote{We are here using Sweedler notation for which $\Delta a = a^{(1)} \otimes a^{(2)} \equiv \sum_i a^{(1)}_i \otimes a^{(2)}_i $, the summation being suppressed.}
\begin{equation*}
[N^{(1)}_j\act x^\mu,N^{(2)}_j\act dx^\nu] = N_j\act[x^\mu,dx^\nu] \ .
\end{equation*}
This is due to the fact that the covariance condition we are here proposing is based necessarily on the whole Lorentz transformation (\ref{Lorentz}) connecting the coordinates of boosted and rotated observers, and this involves, as required for the consistency of the Lorentz action (see Sec.~\ref{sec:covarianceCheck}), boost and rotation parameters that do not commute with coordinates.
    
\section{Four dimensional differential calculus}

To keep the discussion focused on the relativistic aspects of the mathematical construction, we define our four dimensional differential calculus, following the proposal of~\cite{GACagoDandreakMinkTransl} (leading to the characterization of Noether currents in~\cite{GACkNoether}), starting from the description of translations in $\kappa$-Minkowski spacetime ${\cal M}_\kappa$. This is defined by the commutation relations and coproducts
\begin{equation}
[x^0,x^j] = \frac{i}{\kappa} x^j \ , \qquad
[x^j,x^k] = 0 \ ,
\label{kMink}
\end{equation}
\begin{equation*}
\Delta x^\mu = x^\mu \otimes \un + \un + x^\mu \ .
\end{equation*}

To introduce the $\kappa$-Poincar\'e (Hopf-) algebra ${\cal P}_\kappa$ of relativistic symmetries of ${\cal M}_\kappa$, one can start~\cite{GACagoDandreakMinkTransl} from the definition of time-ordered plane waves $:e^{ikx}:$, that serve as base for the (noncommutative) Fourier expansion of functions:
\begin{equation*}
f(x) = \int d\mu(k) \tilde{f}(k) :e^{ikx}:\ , 
\end{equation*}
with $d\mu(k)$ a suitable (invariant) measure in momentum space.
From a geometrical perspective~\cite{KowaNowaFrePLB,KowaNowaFre,kdiscrete}, time-ordered plane waves represent elements of the (Borel) group $AN_3$. From this, dual, perspective, momentum space is a curved manifold, momenta are ``coordinates'' on $AN_3$ (that cover half of de Sitter space), and $x^\mu$ generate (covariant, dual) ``translations'' in momentum space.
Choosing, for instance, the time-to-the-right plane waves
\begin{equation*}
:e^{ikx}: = e^{ik_j x^j} e^{ik_0 x^0} \ ,
\end{equation*}
one can define translations as generated by operators $P_\mu$ whose ``eigenvalues'' are the momenta $k_\mu$:
\begin{equation}
P_\mu \act :e^{ikx}: = k_\mu :e^{ikx}: \ .
\label{Paction}
\end{equation}
It is easy to show that, on products of functions, the action of $P_j$ doesn't satisfy the Leibniz rule. Its deformed action on products of functions is encoded in the non-primitive coproduct
\begin{equation}
\Delta P_j = P_j \otimes \id +  e^{-i P_0/\kappa} \otimes P_j \ .
\label{coproductPj}
\end{equation}
The time translation generator $P_0$ acts with the standard Leibniz rule, and its coproduct is primitive
\begin{equation}
\Delta P_0 = P_0 \otimes \id +  \id \otimes P_0 \ .
\label{coproductP0}
\end{equation}

One can show~\cite{GACagoDandreakMinkTransl} that also for boosts and rotations, the consistency of their action on time-to-the-right plane waves requires their algebra and coalgebra to be $\kappa$-deformed. In particular, identifying the commutation of generators by the commutation of their action on functions on ${\cal M}_\kappa$, i.e. for instance
\begin{equation*}
N_j \act P_\mu \act f(x)- P_\mu \act N_j \act f(x)= [N_j,P_\mu] \act f(x) \ ,                                                                                                                      \end{equation*}
one gets the expressions
\begin{gather}
[P_\mu, P_\nu] = 0 \ , \quad [M_j, P_0] = 0 \ , \quad [N_j, P_0] = i P_j \ , \nonumber \\ 
[N_j, P_k] = i\delta_{jk}\left(\frac{\kappa}{2}\left(1-e^{-2 P_{0}/\kappa}\right)+\frac{1}{2\kappa}\vec{P}^{2}\right)-\frac{i}{\kappa} P_{j}P_{k} \ , \nonumber\\ 
[M_j, P_k] = i \epsilon_{jkl} P_l\ , \quad  [M_j , N_k] = i \epsilon_{jkl} N_l\ , \label{kPoincAlgebra}\\ 
[M_j , M_k] = i \epsilon_{jkl} M_l\ ,  \quad [N_j , N_k] = - i \epsilon_{jkl} M_l\ , \nonumber
\end{gather}
\begin{equation}
\begin{gathered}
\Delta M_{j}=M_{j}\otimes \id + \otimes M_{j}\ ,\\
\Delta N_j = N_{j} \otimes \id +e^{-P_{0}/\kappa}\otimes N_{j}+\frac{1}{\kappa}\epsilon_{jkl}P_{k}\otimes M_{l} \ , 
\end{gathered}
\label{LorentzCoproduct}
\end{equation}
that, together with (\ref{coproductPj}) and (\ref{coproductP0}), define the $\kappa$-Poincar\'e Hopf-algebra in bicrossproduct basis proposed in~\cite{MajidRuegg}.

In order to describe translations, however, the properties of the generators are not enough. 
One has to consider the role of the translation parameters $a^\mu$ that define the infinitesimal translation operator
\begin{equation*}
T = (\id + i a^\mu P_\mu) \triangleright \ ,
\end{equation*}
which, in order to comply with the properties of $\kappa$-Minkowski, turn out to have nontrivial commutation rules with $x^\mu$.
A requirement that one wants to impose on translation parameters, leading to a suitable description of translation symmetries~\cite{GACagoDandreakMinkTransl,GACkNoether}, is that their combination with translation generators restores the Leibniz rule for the operator $i a^\mu P_\mu \act $:
\begin{equation*}
i a^\mu P_\mu \act (f(x) g(x) ) = (i a^\mu P_\mu \act f(x)) g(x) + f(x) i a^\mu P_\mu \act g(x) 
\end{equation*}
Given (\ref{coproductPj}), and asking also that $P_\mu \act a^\nu = 0$ and $[a^\mu,a^\nu] = 0$, this requirement singles out the possible choice of commutation relations to
\begin{equation}
\begin{gathered}
{} [x^0,a^0 ] = [x^j,a^\mu] = 0 \ , \qquad [x^0 , a^j] = \frac{i}{\kappa} a^j \ ,
\end{gathered}
\label{translationComm}
\end{equation}
It turns out that, besides the Leibniz rule, these commutation properties of translation parameters ensure that the map $T \act$ is an algebra isomorphism,
\begin{equation*}
[T\act x^\mu , T\act x^\nu ] =  T\act [x^\mu , x^\nu ]\ ,
\end{equation*}
i.e. translations transform elements of ${\cal M}_\kappa$ in elements of a second copy of ${\cal M}_\kappa$, or, in other words, that for a translated observer the noncommutative spacetime is still defined by (\ref{kMink}).
These properties of $a^\mu$, Leibniz rule and algebra isomorphism, are crucial for the description of translational symmetries for noncommutative field theories in $\kappa$-Minkowski~\cite{GACkNoether}, as they allow to perform Noether analysis  establishing a correspondence between differentiation and functional variation of the fields.
It is also clear that the same properties are exactly the ones of bicovariance, discussed in Sec.~\ref{sec:covariance}, Eqs. (\ref{diffCalcLeibniz}) and (\ref{diffCalcIsomorph}), that a differential calculus on $\kappa$-Minkowski must satisfy. 
We can thus define our differential calculus on ${\cal M}_\kappa$ by the identification
\begin{equation*}
d_T := i a^\mu P_\mu \act\ , \ \text{i.e.} \quad d_Tx^\mu = i a^\nu P_\nu \act x^\mu = a^\mu \ ,
\end{equation*}
where we have used that from the definitions (\ref{Paction}) follows obviously that $P_\mu \act x^\nu = -i \delta_\mu^\nu$.
So that, from (\ref{translationComm}),
\begin{equation}
[x^0,d_Tx^0] = [x^j,d_Tx^\mu] = 0 \ , \quad [x^0 , d_Tx^j] = \frac{i}{\kappa} d_Tx^j \ .
\label{diffTcomm}
\end{equation}
It turns out that this differential calculus, that we have obtained from relativistic considerations involving translational symmetry, coincides with the one found, for $\kappa$-Minkowski spacetime, in~\cite{OecklDiffCalc}, where it is shown moreover to be compatible with the introduction of a translational invariant notion of integration.

\section{Covariance of the four dimensional calculus}
\label{sec:covarianceCheck}

It now comes the main part of this letter, the illustration that the four dimensional calculus induced by translation parameters, is not only bicovariant, but also covariant under Lorentz transformations.
In order to show that, we must first discuss the role of Lorentz parameters in the operator (\ref{Lorentz}). Similarly to what was done for translations, we impose that the boost and rotations parameters combine with the coproducts (\ref{LorentzCoproduct}) of the corresponding $\kappa$-Poincar\'e generators so that their action on products of functions satisfy the Leibniz rule.
This construction was performed in~\cite{GACnopure}, and, again, its physical motivation lies in the fact that with this property one can perform Noether analysis by variational principle on noncommutative field theory.
The resulting commutation relation, asking also that $P_\mu \act \xi^j = P_\mu \act \rho^j = 0$ and $[\xi^j,\rho^k] = 0$, are
\begin{equation}
\begin{gathered}
{} [x^0,\xi^j ] = \frac{i}{\kappa} x^j \ ,\quad [x^j,\xi^k ] = 0\ ,\\
[x^0,\rho^j ] = 0 \ ,\quad [x^j,\rho^k ] = - \frac{i}{\kappa} \epsilon_{jkl}\xi^l\ ,
\end{gathered}
\label{LorentzComm}
\end{equation}
From these commutation relations, one can see~\cite{GACnopure} that boosts and rotations are intertwined, so that one cannot perform a boost without producing a ($\kappa$-suppressed) rotation (a feature denoted as ``no-pure'' boost in~\cite{GACnopure}).
This implies that we have to consider the Lorentz transformation (\ref{Lorentz}) as a whole.

It is noteworthy that (\ref{LorentzComm}) are such that the Lorentz transformed coordinates still satisfy (\ref{kMink}), i.e., $L \act$ is also an algebra isomorphism on ${\cal M}_\kappa$.
In order to show it we first notice that the generators of rotation and boost admit the following representations in terms of $x^\mu$ and $P_\mu$
\begin{equation}
\begin{gathered}
N_j = x^{0}P_{j}+x^{j}\left(\frac{\kappa}{2}( 1-e^{-2 P_{0}/\kappa})+\frac{1}{2\kappa}\vec{P}^{2}\right)\ ,\\
M_j = \epsilon_{jkl} x^k P_l\ .
\end{gathered}
\label{LorentzReps}
\end{equation}
However, their action on ($\kappa$-Minkowski) coordinates is still the standard action~\footnote{\label{foot} Notice that $P_\mu^2 \act x^\nu = P_\mu \act P_\mu \act x^\nu = P_\mu \act -i\delta_\mu^\nu = 0$.}
\begin{equation}
\begin{gathered}
N_j \act x^0 =  - i x^j \ , \quad N_j \act x^k =  - i \delta_{jk} x^0 \ , \\
M_j \act x^0 = 0 \ , \quad M_j \act x^k = i \epsilon_{jkl} x^l \ .
\end{gathered}
\label{LorentzActx}
\end{equation}
Using the last expressions with (\ref{LorentzComm}) and (\ref{kMink}), one finds easily that
\begin{equation*}
[{x'}^0,{x'}^j] = \frac{i}{\kappa} {x'}^j \ , \qquad
[{x'}^j,{x'}^k] = 0 \ ,
\end{equation*}
where
\begin{equation*}
{x'}^\mu = L\act x^\mu = (\id + i \xi^j N_j + i \rho^j M_j)\act x^\mu. 
\end{equation*}

We can now check the Lorentz covariance of the differential calculus.
The first step is to check the action of Lorentz on differentials~(\ref{LorentzActDiff}). Again, we rely on a relativistic argument to see how the translation parameters transform under Lorentz.
The key is to consider the sequence of a Lorentz transformation followed by a translation, and its opposite.
Translations must take into account of length contraction and time dilation (or simply of rotations), so that, when comparing translations before or after Lorentz transformations, the parameters must change accordingly.
If we assume that the only change is in the translation parameters (and not in Lorentz ones), this change will be the action we are looking for.
We thus want to find ${a'}^\mu(a)$ such that
\begin{equation*}
L \act T(a')\act f(x) = T(a)\act L \act f(x) \ .
\end{equation*}
By expanding the actions we get
\begin{equation*}
{a'}^\mu P_\mu \act f(x) = (a^\mu P_\mu - i\xi^j a^\mu [N_j , P_\mu] - i\rho^j a^\mu [R_j , P_\mu] ) \act f(x) \ ,
\end{equation*}
where again, the parameters are assumed to commute between themselves, $[\xi^j,a^\mu]=[\rho^j,a^\mu]=0$.
From the commutation relations (\ref{kPoincAlgebra}), applying last relations to $f(x) = x^\mu$, one finds the standard transformations (see footnote~\ref{foot})
\begin{equation*}
{a'}^\mu = a^\mu + \delta^\mu_0 \xi^j a^j + \delta^\mu_j (\xi^j a^0 + \epsilon_{jkl} \rho^k a^l )
\end{equation*}
Comparing last equation with (\ref{LorentzActx}), it follows that the Lorentz action~(\ref{LorentzActDiff}) on differentials $d_Tx^\mu = a^\mu$ is satisfied:
\begin{equation*}
\begin{split}
& L\act d_Tx^\mu = d_T L\act x^\mu \\ &
= d_T x^\mu + \delta^\mu_0 \xi^j d_T x^j + \delta^\mu_j (\xi^j d_T x^0 + \epsilon_{jkl} \rho^k d_T x^l ) \ .
\end{split}
\end{equation*}

Finally, we have to prove that (\ref{covariance}) is satisfied.
With the last result at hand, we have that
\begin{equation*}
\begin{split}
& {x'}^\mu + d_T{x'}^\mu = L\act (x^\mu + d_Tx^\mu) \\ &
= x^\mu + \delta^\mu_0 \xi^j x^j + \delta^\mu_j (\xi^j x^0 + \epsilon_{jkl} \rho^k x^l ) \\ &
+ d_T x^\mu + \delta^\mu_0 \xi^j d_T x^j + \delta^\mu_j (\xi^j d_T x^0 + \epsilon_{jkl} \rho^k d_T x^l )  \ . 
\end{split}
\end{equation*}
It is now just a matter of calculation to check that, using (\ref{kMink}), (\ref{diffTcomm}) and (\ref{LorentzComm}), it follows
\begin{equation}
\begin{gathered}
{} [{x'}^0 + d_T{x'}^0, {x'}^j + d_T{x'}^j] = \frac{i}{\kappa} ({x'}^j + d_T{x'}^j ) \ , \\
[{x'}^j + d_T{x'}^j,{x'}^k + d_T{x'}^k] = 0 \ .
\label{dTcovariance}
\end{gathered}
\end{equation}
This concludes our proof of the Lorentz covariance of the differential calculus (\ref{diffTcomm}).

\section{Extension to the whole $\kappa$-Poincar\'e algebra}
\label{sec:extension}

Since the commutation rules (\ref{LorentzComm}) of the Lorentz parameters are such that $L\act = (\id + i \xi^j N_j + i \rho^j R_j) \act $ is an algebra isomorphism, with $(i \xi^j N_j + i \rho^j R_j)\act$ satisfying the Leibniz rule, according to the discussion of Sec.~\ref{sec:covariance}, we may try to define a second bicovariant differential calculus $d_L$ on ${\cal M}_\kappa$ associated to Lorentz transformation as
\begin{equation*}
L\act := \id +d_L \ , \qquad d_Lx^\mu = (i \xi^j N_j + i \rho^j R_j)\act x^\mu \ .
\end{equation*}
From (\ref{LorentzActx}), (\ref{kMink}) and (\ref{LorentzComm}), one finds that the these differentials satisfy the commutation relations
\begin{equation*}
\begin{gathered}
{} [x^0, d_Lx^0] = 2\frac{i}{\kappa}d_{L}x^{0} \ , \quad 
\left[x^{0},d_{L}x^{j}\right] = \frac{i}{\kappa} d_{L}x^{j} \ , \\
\left[x^{j},d_{L}x^{0}\right] = 0\ , \quad
\left[x^{j},d_{L}x^{k}\right] = -\frac{i}{\kappa}\delta_{jk}d_{L}x^{0}
\end{gathered}
\end{equation*}
This calculus is bicovariant, and it is obviously also Lorentz covariant (since it is generated by Lorentz transformations), and thus is another example of a four dimensional bicovariant calculus on ${\cal M}_\kappa$.

If we want to rewrite the condition of Lorentz covariance~(\ref{dTcovariance}) of $d_T$ in terms of this second calculus $d_L$, we find that the condition of covariance becomes a condition of compatibility between the two calculi.
For a generic noncommutative spacetime (of Lie algebra type) (\ref{noncommLie}), the covariance condition becomes
\begin{equation*}
\begin{split}
& [x^\mu + d_Lx^\mu + d_Tx^\mu + d_Td_Lx^\mu, x^\nu + d_Lx^\nu + d_Tx^\nu + d_Td_Lx^\nu] \\ & = 
\gamma^{\mu\nu}_\rho(x^\nu + d_Lx^\nu + d_Tx^\nu + d_Td_Lx^\nu) \ ,
\end{split}
\end{equation*}
that can be rewritten, using (\ref{noncommLie}), as
\begin{equation*}
\begin{split}
& [d_Tx^\mu,d_Lx^\nu] + [d_Lx^\mu,d_Tx^\nu] + [x^\mu,d_Td_Lx^\nu] + [d_Td_Lx^\mu,x^\nu] \\
& = i \gamma^{\mu\nu}_\rho d_Td_Lx^\rho = d_Td_L [x^\mu,x^\nu] \ .
\end{split}
\end{equation*}
These relations show the close relationship between the relativistic structure of the $\kappa$-Poincar\'e framework and the properties of the differential calculus.

\section{Scalarity of noncommutative plane wave}

The presence of noncommutative parameters associated to Lorentz transformations may seem awkward at first sight.
However, there is at least another very good reason for their necessity.
In analogy with classical field theory, we would like a scalar function on ${\cal M}_\kappa$ to transform, under an infinitesimal Lorentz transformations, as
\begin{equation}
 f(x') = L\act f(x) \ .
\label{scalarf}
\end{equation}
By Fourier transform, we can formulate this equation in terms of (noncommutative) plane waves. Then, the l.h.s. implies (these equations must be understood to be valid at first order in the transformation parameters)
\begin{equation*}
:e^{ikx'}: = e^{ik_j(x^j + \xi^j x^0 + \epsilon_{jkl} \rho_k x^l)} e^{ik_0(x^0 + \xi^j x^j)} \ .
\end{equation*}
Using the commutation relations (\ref{kMink}) and (\ref{LorentzComm}), we can bring out (to the left) of the exponentials the terms containing the parameters in the last expression. After some calculation, one gets, at first order in the parameters, that the expression becomes
\begin{equation*}
\begin{split}
 :e^{ikx'}: = &  \Big(1+i\xi^{j} \big(x^{0}k_{j}+x^{j} \big(\frac{\kappa}{2}( 1-e^{-2 k_{0}/\kappa}) +\frac{1}{2\kappa}\vec{k}^{2} \big) \big) \\ & 
+i\rho^{j}\epsilon_{jkl}x^{k}k_{l}\Big) :e^{ikx}: \ .
\end{split}
\end{equation*}
But, comparing it with (\ref{LorentzReps}), it follows that it is exactly
\begin{equation*}
(\id +i\xi^{j}N_{j}+i\rho^{j}R_{j})\act :e^{ikx}: = L\act :e^{ikx}:\ ,
\end{equation*}
so that (\ref{scalarf}) holds. It is crucial, in the calculation, the role of the noncommutative parameters in order to reproduce the representations (\ref{LorentzReps}). We conclude that the introduction of the noncommutative parameters (\ref{LorentzComm}) is necessary to define a scalar function in $\kappa$-Minkowski.

\section{Pregeometric representation}

In~\cite{pregeometry} we have proposed a representation of both $\kappa$-Minkowski coordinates and $\kappa$-Poincar\'e symmetries in (1+1)D, including the noncommutative transformation parameters, in terms of a a standard (two-dimensional) Heisenberg algebra of operators. Such a representation is called in the literature ``pregeometric''.
We want here to extend this representation to the four dimensional case. This will shed some light on some features of the noncommutative calculus.
Given the (four dimensional) Heisenberg algebra
\begin{equation*}
[\pi_\mu, q^\nu] =  - i\delta_\mu^\nu \ ,
\end{equation*}
${\cal M}_\kappa$ coordinates can be represented as
\begin{equation}
x^0 = q^0 \ , \qquad x^j = e^{\pi_0/\kappa} q^j \ .
\label{kMinkPre}
\end{equation}
Translation generators are given by~\cite{pregeometry}
\begin{equation}
P_0 \act = [\pi_0 , \cdot ] \ , \qquad P_j \act = e^{-\pi_0/\kappa}[\pi_j , \cdot ] \ ,
\label{transPre}
\end{equation}
while translation parameters are
\begin{equation}
a^0 = \tilde{a}^0 \ , \qquad a^j = \tilde{a}^j e^{\pi_0/\kappa} \ ,
\label{transParPre}
\end{equation}
with $\tilde{a}^\mu$ real (commutative) numbers.
It follows that the action of the translation operator is
\begin{equation}
ia^\mu P_\mu \act = i[\pi_\mu,\cdot] \ , 
\label{transLeibniz}
\end{equation}
that manifestly satisfies the Leibniz rule, since it acts by commutation.

Boost and rotation generators are given by combining (\ref{LorentzReps}) and (\ref{kMinkPre}).
It is easy to find the pregeometric representation also for boost and rotation generators (\ref{LorentzComm}). They are
\begin{equation}
\xi^j = \tilde{\xi}^j e^{\pi _0 /\kappa}\ , \qquad \rho^j = \tilde{\rho}^j - \frac{1}{\kappa}\epsilon_{jkl} \tilde{\xi}^k \pi_l  \ ,
\label{LorentzParPre}
\end{equation}
where $\tilde{\xi}^j$ and $\tilde{\rho}^j$ are real numbers.
From the last expressions, it is apparent how rotations and boosts are intertwined so that if boosts are different from zero, the parameter of rotation is also switched on. This is the manifestation of what was denoted in~\cite{GACnopure} as no-pure boost transformation.
Considering the representation of the Lorentz generators (\ref{LorentzReps}), from the pregeometric representations (\ref{kMinkPre}), (\ref{transPre}) and (\ref{LorentzParPre}), one can show that the total Lorentz operator is
\begin{equation}
(i \xi^jN_j + i \rho^jM_j) \act =  i[\tilde{\xi}^j \eta_j + \tilde{\rho}^j \mu_j ,\cdot] \ ,
\label{LorentzOpPre}
\end{equation}
where
\begin{equation*}
\begin{gathered}
\eta_{j}=\pi_{j}q^{0}+\left(\left(\frac{\kappa}{2}(e^{2\pi_{0}/\kappa}-1)-\frac{1}{2\kappa}\vec{\pi}^{2}\right)\delta_{jk} + \frac{1}{\kappa}\pi_{j}\pi_{k}\right)q^{k}\ ,\\
\mu_{j}=\epsilon_{jkl}\pi_{l}q^{k}\ .
\end{gathered}
\end{equation*}
Like for translations, Eq. (\ref{LorentzOpPre}) tell us that the Lorentz operator, since it acts by commutator, satisfies the Leibniz rule, as expected.

\section{Aside on the five dimensional calculus}

So far we didn't mention the five dimensional calculus proposed in~\cite{Sitarz5dCalc}. Even if, with a covariant four dimensional calculus at our disposal, one can avoid the introduction of a fifth translation parameter, whose physical interpretation may be challenging, however, it is still an open question, having two covariant calculi at our disposal, what is the relation between them.
While we postpone a more detailed investigation of this question to future studies, we want to conclude this letter with a few considerations that may be useful also in this respect.

The first consideration is of geometrical nature. It is well known~\cite{LukKosMasSitkField,KowaNowaFrePLB,KowaNowaFre,kdiscrete}, that the five dimensional calculus $d_{5D}$ proposed in~\cite{Sitarz5dCalc} can be understood to correspond to spacetime transformations
\begin{equation*}
f(x')=(1+iA^\mu\bar{P}_\mu + i A^4 (\bar{P}_4 - \kappa ))\act f(x) \ , 
\end{equation*}
where the parameters $A^a\equiv d_{5D}x^a$ ($a=0,1,2,3,4$), with $A^4 \equiv d_{5D}x^4$ the fifth differential form, reproduce the commutation rules of the calculus
\begin{equation}
\begin{gathered}
\left[x^{0},A^{0}\right]=-\tfrac{i}{\kappa} A^{4}\ ,\quad\left[x^{0},A^{4}\right]=-\tfrac{i}{\kappa} A^{0}\ ,\\
\left[x^{j},A^{0}\right]=\left[x^{j},A^{4}\right]=- \tfrac{i}{\kappa} A^{j}\ ,\\
\left[x^{0},A^{j}\right]=0\ ,\quad \left[x^{j},A^{k}\right]= \tfrac{i}{\kappa} \left(A^{4}-A^{0}\right)\delta_{jk}\ ,
\end{gathered}
\label{5DcalcComm}
\end{equation}
and with generators $\bar{P}_a$ that coincide with ``coordinates'' in five dimensional flat, Minkowskian, momentum space given by the embedding of the de Sitter hyperboloid (or better, of the $AN_3$ manifold) on 5D Minkowski space
\begin{equation}
\begin{gathered}\bar{P}_{0}=\kappa\sinh\left(P_{0}/\kappa\right)+\tfrac{1}{2\kappa}e^{P_{0}/\kappa}\vec{P}^{2}\ ,\\
\bar{P}_{4}=\tfrac{1}{\kappa}\cosh\left(P_{0}/\kappa\right)-\tfrac{1}{2\kappa}e^{ P_{0}/\kappa}\vec{P}^{2}\ .
\end{gathered}
\quad\bar{P}_{j}=e^{ P_{0}/\kappa}P_{j}\ ,\label{embedding}
\end{equation}
If we regard translations as describing a ``rigid'' shift of spacetime coordinates in the direction of the generating momenta, then we can argue that the transformations corresponding to the two calculi are of different nature: for the four dimensional calculus they correspond to shifts in the direction of momenta $P_\mu$ ``on the de Sitter (momentum space) hyperboloid''; while for the five dimensional calculus they correspond to shifts of the $x^\mu$ (and functions of $x^\mu$) in the direction of flat momenta $\bar{P}_a = (\bar{P}_\mu, \bar{P}_4)$ characterizing the surrounding five dimensional Minkowskian environment in which the de Sitter hyperboloid is embedded.
From this perspective it is also clear why the translations associated to the 5D calculus have generators that satisfy the standard, flat, Poincar\'e commutators with boosts and rotations: the 5D embedding momenta generators live in flat Minkowskian 5D space, and thus satisfy standard special relativistic commutation rules with Lorentz generators.
However, besides the possible implications for phenomenology that we mentioned in the introduction, this choice of translations, in the direction of the surrounding (embedding) flat momenta, instead of the momenta on the hyperboloid, seems rather ad hoc.
Thus, while the 5D calculus has a mathematical relevance on its own, and may have as well physical applications, we feel that, if we want to apply the calculus to the description of spacetime translations for $\kappa$-Poincar\'e/$\kappa$-Minkowski framework, whose momenta are described as the $AN_3$ group manifold, and coordinates as the ``generators of translations in (de Sitter) momentum space'', it may be preferable to adopt the four dimensional calculus and its geometrical interpretation.

The second consideration regards the pregeometric representation of the 5D calculus. It is possible indeed to show that also the 5D calculus admits such a representation, so that the commutators (\ref{5DcalcComm}) are satisfied:
\begin{equation}
\begin{gathered}A^{0}=e^{\pi_{0}/\kappa}\alpha^{0} - \tfrac{1}{\kappa}e^{-\pi_{0}/\kappa}\pi_{j}\left(\alpha^{j}-\tfrac{1}{\kappa}\alpha^{0}\pi_{j}\right)\ ,\\
A^{j}=\alpha^{j}- \tfrac{2}{\kappa} \alpha^{0}\pi_{j}\ ,\\
A^{4}=-e^{\pi_{0}/\kappa}\alpha^{0}- \tfrac{1}{\kappa} e^{-\pi_{0}/\kappa} \pi_{j}\left(\alpha^{j}-\tfrac{1}{\kappa}\alpha^{0}\pi_{j}\right)\ ,
\end{gathered}
\label{5DtransParPre}
\end{equation}
with $\alpha^\mu$ numerical, commutative, constants.
It follows by careful computation, that the transformation operator associated to this calculus, is
\begin{equation*}
\begin{split}
& d_{5D} \equiv (iA^\mu\bar{P}_\mu + i A^4 (\bar{P}_4 - \kappa )) \act \\ & 
=  \Big[\alpha^{0} \Big(\kappa (e^{\pi_{0}/\kappa}-1 )- \tfrac{1}{\kappa} e^{-\pi_{0}/\kappa}\vec{\pi}^{2} \Big)+\alpha^{j}\pi_{j}e^{-\pi_{0}/\kappa},\cdot \Big]
\end{split}
\end{equation*}
The last relation shows that the operator satisfies the Leibniz rule.
On the other side, if we compare it with (\ref{transLeibniz}), we see that the two transformations are substantially  different.
Moreover, relations (\ref{5DtransParPre}) manifest another feature: the translations generated by the 5D calculus, differently from the four dimensional case~(\ref{transParPre}), are intertwined together so that you cannot switch off the 5th extra parameter, in a way similar to the no-pure transformations of the previous section.

\section{Conclusions}

Motivated by the implications that a differential calculus has for the physical interpretation of the theory, we revisited the argument~\cite{Sitarz5dCalc} against the (Lorentz) covariance of a four dimensional differential calculus on $\kappa$-Minkowski noncommutative spacetime, with $\kappa$-Poincar\'e relativistic symmetries.
We found in particular that the notion of (Lorentz) covariance introduced in~\cite{Sitarz5dCalc} does not take into account of the properties of the noncommutative Lorentz parameters of transformations~\cite{GACnopure}.
These properties, however, introduced in~\cite{GACnopure} with the purpose of performing Noether analysis on ${\cal M}_\kappa$ noncommutative field theory, turn out to be fundamental for the consistence of the whole setup of $\kappa$-Poincar\'e relativistic transformations, as we have here illustrated.
Once taking correctly into account of the properties of the noncommutative Lorentz sector parameters, the four dimensional calculus introduced in~\cite{OecklDiffCalc}, which is at the basis of several studies aiming to investigate the Planck-scale phenomenological implications of theories with $\kappa$-deformed symmetries, turns out to be fully Lorentz covariant.
The construction presented in this letter can be extended to the whole $\kappa$-Poincar\'e algebra of relativistic transformations, where the parameters of the Lorentz sector describe a second four dimensional calculus on ${\cal M}_\kappa$ (sec.~\ref{sec:extension}).
We have also given the pregeometric representations of the calculi, that help to clarify some of their features, and outlined some considerations for a comparison between the 4D and 5D calculi.

\section*{Acknowledgements}

I would like to thank Valerio Astuti for helpful discussions during the writing of this manuscript.
This work was supported by funds provided by the National Science Center, project number  2019/33/B/ST2/00050.

\end{document}